\documentclass[aps,prb,twocolumn,superscriptaddress]{revtex4}
\makeatletter

\newcommand{\Rmnum}[1]{\expandafter\@slowromancap\romannumeral #1@}
\makeatother

\usepackage{eurosym}
\usepackage{amsfonts}
\usepackage{amssymb}
\usepackage{amsmath}
\usepackage{graphicx}
\usepackage{color}
\begin{document}

\title{Strain-induced enhancement of thermoelectric performance in a ZrS$_{2}$ monolayer}

\author{H. Y. Lv}

\affiliation{Key Laboratory of Materials Physics, Institute of Solid State Physics, Chinese Academy of Sciences, Hefei 230031, People's Republic of China}

\author{W. J. Lu}
\email[Corresponding author: ]{wjlu@issp.ac.cn}
\affiliation{Key Laboratory of Materials Physics, Institute of Solid State Physics, Chinese Academy of Sciences, Hefei 230031, People's Republic of China}

\author{D. F. Shao}

\affiliation{Key Laboratory of Materials Physics, Institute of Solid State Physics, Chinese Academy of Sciences, Hefei 230031, People's Republic of China}

\author{H. Y. Lu}

\affiliation{School of Physics and Electronic Information, Huaibei Normal University, Huaibei 235000, People's Republic of China}

\author{Y. P. Sun}
\email[Corresponding author: ]{ypsun@issp.ac.cn}
\affiliation{Key Laboratory of Materials Physics, Institute of Solid State Physics, Chinese Academy of Sciences, Hefei 230031, People's Republic of China}
\affiliation{High Magnetic Field Laboratory, Chinese Academy of Sciences, Hefei 230031, People's Republic of China}
\affiliation{Collaborative Innovation Center of Advanced Microstructures, Nanjing University, Nanjing, 210093, People's Republic of China}

\makeatletter


\begin{abstract}
The increase of a thermoelectric material's figure of merit ($ZT$ value) is limited by the interplay of the transport coefficients. Here we report the greatly enhanced thermoelectric performance of a ZrS$_{2}$ monolayer by the biaxial tensile strain, due to the simultaneous increase of the Seebeck coefficient and decrease of the thermal conductivity. Based on the first-principles calculations combined with the Boltzmann transport theory, we predict the band gap of the ZrS$_{2}$ monolayer can be effectively engineered by the strain and the Seebeck coefficient is significantly increased. The thermal conductivity is reduced by the applied tensile strain due to the phonon softening. At the strain of 6\%, the maximal $ZT$ value of 2.4 is obtained for the $p$-type doped ZrS$_{2}$ monolayer at 300 K, which is 4.3 times larger than that of the unstrained system.

\end{abstract}
\maketitle

Thermoelectric (TE) devices can directly convert the heat energy to electricity and vice versa, and thus have promising applications in solid-state cooling and power generation. The efficiency of a TE device is determined by the TE material's dimensionless figure of merit $ZT=S^2{\sigma}T/(\kappa_e+\kappa_p)$, where $S$, $\sigma$, $T$, $\kappa_e$ and $\kappa_p$ are the Seebeck coefficient, electrical conductivity, absolute temperature, electronic and lattice thermal conductivities, respectively. Because the transport coefficients ($S$, $\sigma$, $\kappa_e$ and $\kappa_p$) are strongly coupled to each other, none of them can be independently tuned to largely enhance the TE performance.

It was theoretically proposed that one- (1D) or two-dimensional (2D) materials could have much larger $ZT$ values than their bulk counterparts.\cite{MS1,MS2} In recent years, with the rapid pace of progress in nanotechnologies, a large variety of low-dimensional materials beyond graphene have been successfully prepared.\cite{synthesize1,synthesize3}  Among them, single layers of transition metal dichalcogenides (TMDCs) have received a lot of attention because some of them, such as MoS$_{2}$, are semiconductors with sizable direct band gaps, making them promising candidates for applications in field effect transistors and optoelectronic devices. In addition, MX$_{2}$ (M=Mo, W; X=S, Se) monolayers were reported to have much improved TE performance,\cite{NanoLett-Buscema-2013,JAP-Huang-2013,JCP-Wickramaratne-2014,APL-Babaei-2014} opening up a new opportunity of TMDCs monolayers in the TE field.

A ZrS$_{2}$ monolayer is another typical 2D TMDC and has been successfully synthesized experimentally.\cite{synthesize4} The thermal conductivity of the ZrS$_{2}$ monolayer is much lower than those of MX$_{2}$ (M=Mo, W; X=S, Se) monolayers,\cite{APL-phonon} which is desirable for the TE application. However, the TE performance of the ZrS$_{2}$ monolayer has not been explored. Recently, it was reported that the electronic structure of the ZrS$_{2}$ monolayer can be tuned by external strains.\cite{RSCAdv-Li-2014} In previous reports, strain has proven to be a flexible and effective method to tune the electronic,\cite{ACSNano-Ni-2008,NanoLett-He-2013,NanoLett-Fei-2014,PRB-Lv-2014} phonon,\cite{PRB-Hu-2013,Nanotech-Zhu-2015} and thus the TE properties of 2D systems. For example, the Seebeck coefficient of phosphorene can be greatly enhanced due to strain-induced band convergence.\cite{PRB-Lv-2014} The thermal conductivity can either be increased\cite{PRB-Hu-2013} or reduced\cite{Nanotech-Zhu-2015} by the applied strains, depending on the particular materials. It is thus interesting to investigate how the TE performance of the ZrS$_{2}$ monolayer will be influenced by the external strain. In this work, by using the first-principles calculations combined with the Boltzmann transport theory, we predict that the TE performance of the ZrS$_{2}$ monolayer can be largely enhanced by the biaxial tensile strain, due to the simultaneous increase of the Seebeck coefficient and decrease of the thermal conductivity.

\begin{figure}
\includegraphics[width=1.0\columnwidth]{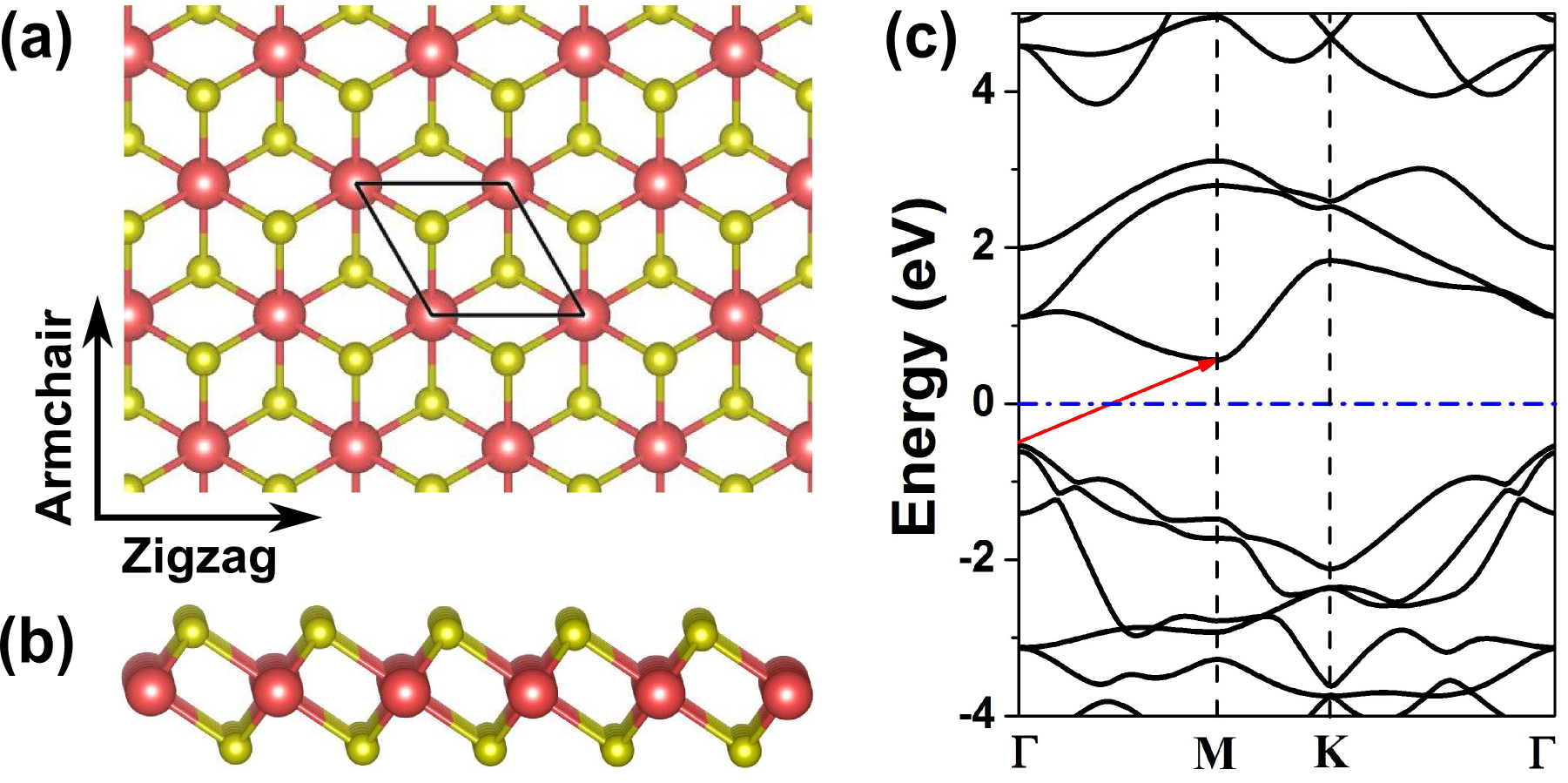}\caption{\label{fig1-structure and energy band}(a) Top and (b) side views of ZrS$_{2}$ monolayer. (c) Band structure of the unstrained ZrS$_{2}$ monolayer. The black line in (a) denotes the primitive cell used in our calculations. The red arrow points from the valence band maximum to the conduction band minimum.}
\end{figure}

The structural and electronic properties of the ZrS$_{2}$ monolayer were calculated within the framework of the density functional theory (DFT),\cite{PR-DFT} as implemented in the ABINIT code.\cite{ABINIT1,ABINIT2,ABINIT3} The projector augmented-wave (PAW) pseudopotential approach were used. The exchange-correlation potential was in form of the Perdew-Burke-Ernzerhof (PBE) expression\cite{PBE} of the generalized-gradient approximation (GGA). The spin-orbital coupling was included in the calculations of electronic properties. The cutoff energy was set to be 600 eV. For each monolayer, a vacuum region of 15 {\AA} was added so that the interactions between the monolayer and its period image can be neglected. The Brillouin zones were sampled with $7\times7\times1$ and $10\times10\times1$ Monkhorst-Pack $k$ meshes for the structural relaxation and electronic structure calculations, respectively. The electronic transport coefficients are derived from the electronic structure based on the semiclassical Boltzmann transport theory, as implemented in the BOLTZTRAP code.\cite{Boltzmann} Doping is treated within the rigid band picture.\cite{rigid-band} The electronic thermal conductivity is calculated by the Wiedemann-Franz law $\kappa_e=L\sigma T$, where $L$ is the Lorenz number. Here we use a Lorenz number of $1.5\times10^{-8}$ W$\Omega$/K$^2$.\cite{Lorenz}

\begin{figure}
\includegraphics[width=0.8\columnwidth]{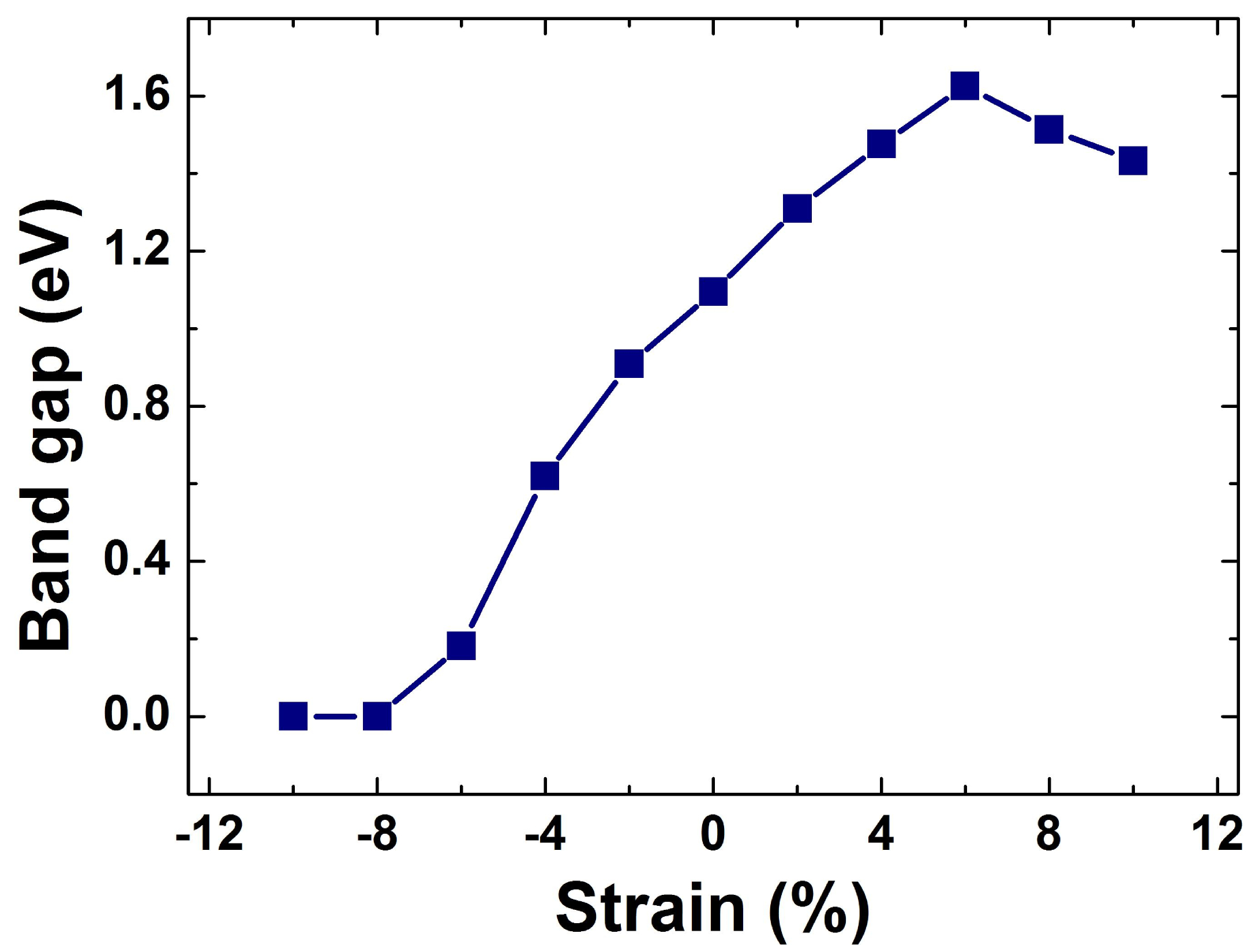}\caption{\label{fig2-band gap}Calculated band gap of ZrS$_{2}$ monolayer as a function of the applied biaxial strain.}
\end{figure}

In the phonon calculations, the force constant matrices were calculated by density functional perturbation theory (DFPT) as implemented in the VASP code.\cite{vasp1,vasp2,vasp3} A $5\times5\times1$ superlattice was used. Phonon frequencies were obtained by the PHONOPY program.\cite{phonopy} The lattice thermal conductivity was calculated by solving the phonon Boltzmann transport equation within the relaxation time approximation, as implemented in the ShengBTE code.\cite{ShengBTE} The second order harmonic and third order anharmonic interatomic force constants (IFCs) were calculated by using $5\times5\times1$ supercell with $2\times2\times1$ Monkhorst-Pack $k$ meshes and $4\times4\times1$ supercell with $\Gamma$ point, respectively.  The interactions up to third-nearest neighbors were considered when calculating the third order IFCs.

The top and side views of the ZrS$_{2}$ monolayer are shown in Figs. 1(a) and (b), respectively. After full relaxation, the lattice parameters are $a=b=3.676$ {\AA}, slightly larger than the experimental values of the ZrS$_{2}$ bulk.\cite{bulk} The unstrained ZrS$_{2}$ monolayer is semiconducting, with the valence band maximum (VBM) and conduction band minimum (CBM) located at $\Gamma$ and $M$ points, respectively, as shown in Fig. 1(c). The indirect band gap is calculated to be 1.10 eV, in good agreement with the previous report.\cite{RSCAdv-Li-2014}

Next, we will investigate the strain effect on the electronic structure of the ZrS$_{2}$ monolayer. Here the in-plane biaxial strain is considered, which is defined as $\varepsilon=(a-a_0)/a_0$, where $a$ and $a_0$ are the in-plane lattice constants of the strained and unstrained monolayers, respectively. The band gap as a function of the applied strain is shown in Fig. 2. When the tensile strain is applied, the system remains an indirect-band-gap semiconductor up to the strain of 10\%. The band gap first increases as increasing the strain and reaches a maximal value of 1.63 eV at the strain of 6\%. When the strain is further increased, the band gap then decreases. When the the strain is compressive, the band gap decreases monotonically as increasing the strain. At the strain of $-8\%$, the ZrS$_{2}$ monolayer tends to be a metal and the semiconductor-to-metal transition takes place.

\begin{figure}
\includegraphics[width=0.8\columnwidth]{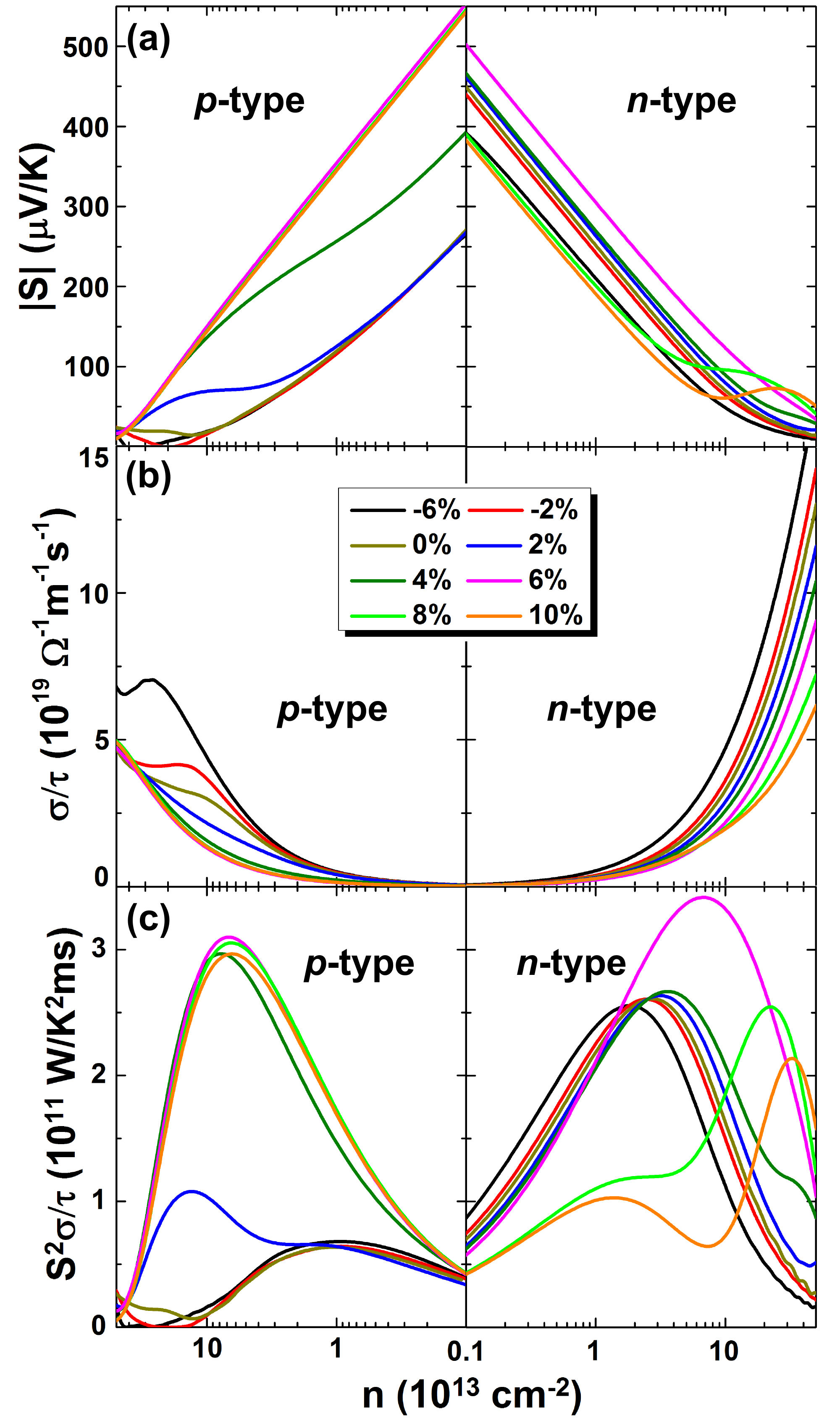}\caption{\label{fig3-electronic transport}(a) Absolute values of the Seebeck coefficient, (b) electrical conductivity and (c) power factor of ZrS$_{2}$ monolayer as a function of the carrier concentration under different biaxial strains.}
\end{figure}

\begin{table*}

\caption{\label{Table1-relaxation time}Effective mass ($m^*$), average effective mass ($m_d$), elastic modulus $C$, DP constant $E_1$, carrier mobility ($\mu$), relaxation time ($\tau$) at 300 K in the zigzag and armchair directions of the unstrained and 6\% strained ZrS$_{2}$ monolayers.}

\begin{tabular}{ccccccccccccccccc}
\hline
\hline
~&~&~&~&~&~& $m^*$ &~& $m_d$ &~& $C$ &~& $E_1$ &~& $\mu$ &~& $\tau$ \tabularnewline
~&~&~&~&~&~& ($m_e$) &~& ($m_e$) &~& (eV/({\AA}$^2$)) &~& (eV) &~& (cm$^2$V$^{-1}$s$^{-1}$) &~& ($10^{-13}$ s) \tabularnewline
\hline
Unstrained &~& Zigzag &~& Electron &~& 0.28 &~& 0.74 &~& 6.36 &~& $-$3.17 &~& 1045.3 &~& 1.66 \tabularnewline
~&~&~&~&  Hole &~& $-$0.27 &~& 0.26 &~& 6.36 &~& $-$6.67 &~& 695.8 &~& 1.07  \tabularnewline
~&~& Armchair &~& Electron &~& 1.97 &~& 0.74 &~& 6.34 &~& $-$3.99 &~& 93.3 &~& 1.04 \tabularnewline
~&~&~&~& Hole &~& $-$0.26 &~& 0.26 &~& 6.34 &~& $-$6.46 &~& 769.2 &~& 1.14 \tabularnewline
\hline
Strained &~& Zigzag &~& Electron &~& 0.43 &~& 1.13 &~& 6.36 &~& $-$4.02 &~& 276.7 &~& 0.68 \tabularnewline
~&~&~&~&  Hole &~& $-$0.72 &~& 0.80 &~& 6.36 &~& $-$3.61 &~& 289.3 &~& 1.18 \tabularnewline
~&~& Armchair &~& Electron &~& 2.97 &~& 1.13 &~& 6.34 &~& $-$3.73 &~& 46.5 &~& 0.79 \tabularnewline
~&~&~&~& Hole &~& $-$0.90 &~& 0.80 &~& 6.34 &~& $-$3.66 &~& 225.3 &~& 1.15 \tabularnewline
\hline
\hline
\end{tabular}

\end{table*}

Based on the calculated electronic structure, we are able to evaluate the electronic transport coefficients by using the semiclassical Boltzmann transport theory and rigid-band model. Within this method, the Seebeck coefficient $S$ can be calculated independent of the relaxation time $\tau$, however, the electrical conductivity $\sigma$ is calculated with $\tau$ inserted as a parameter, that is, what we obtain is $\sigma/\tau$. In Fig. 3, we plot the calculated electronic transport coefficients of the ZrS$_{2}$ monolayer as a function of the carrier concentration under different strains. The results for the zigzag and armchair directions are the same (due to the high structural symmetry of the ZrS$_{2}$ monolayer), so the transporting directions are not distinguished here. From Fig. 3(a), we can see that the compressive strains have little impact on the Seebeck coefficient of the $p$-type doped system while reduce the Seebeck coefficient of the $n$-type doped one. For the tensile strain, the maximal absolute values of the Seebeck coefficients of both the $p$- and $n$-type doped systems first increase as increasing the strain, reach the maximum at the strain of 6\% and then decrease when the strain is further increased. If we notice the strain-dependence of the band gap (see Fig. 2), we can see that such a trend of the Seebeck coefficient coincides with that of the band gap. Therefore, the Seebeck coefficient is efficiently tuned by the strain $via$ the band-gap engineering. The greatly increased Seebeck coefficient of the ZrS$_{2}$ monolayer is very favorable for the TE application.

Figure 3(b) shows the electrical conductivity $\sigma/\tau$ as a function of the carrier concentration. When the doping is not very high ($n<10^{14} {\mbox{cm}}^{-2}$), $\sigma/\tau$ generally decreases as increasing the band gap, which is in contrast with the tendency of the Seebeck coefficient. The decrease of the electrical conductivity is detrimental to the TE performance. Whether the applied strain will benefit the electronic transport properties or not will be determined by the competition of the two factors. In Fig. 3(c), we plot the power factors of the ZrS$_{2}$ monolayer under different strains. For the tensile strain, the maximal power factor generally increases as increasing the band gap, indicating that the increase of the Seebeck coefficient compensates the negative effect from the electrical conductivity. For the $p$-type doped ZrS$_{2}$ monolayer, the power factor is significantly increased by the applied tensile strain. Compared with the unstrained system, in which the power factor of the $n$-type doping is much larger than that of the $p$-type one, the tensile strain makes the power factors of the $p$- and $n$-type doped systems more balanced.

\begin{figure*}
\includegraphics[width=1.5\columnwidth]{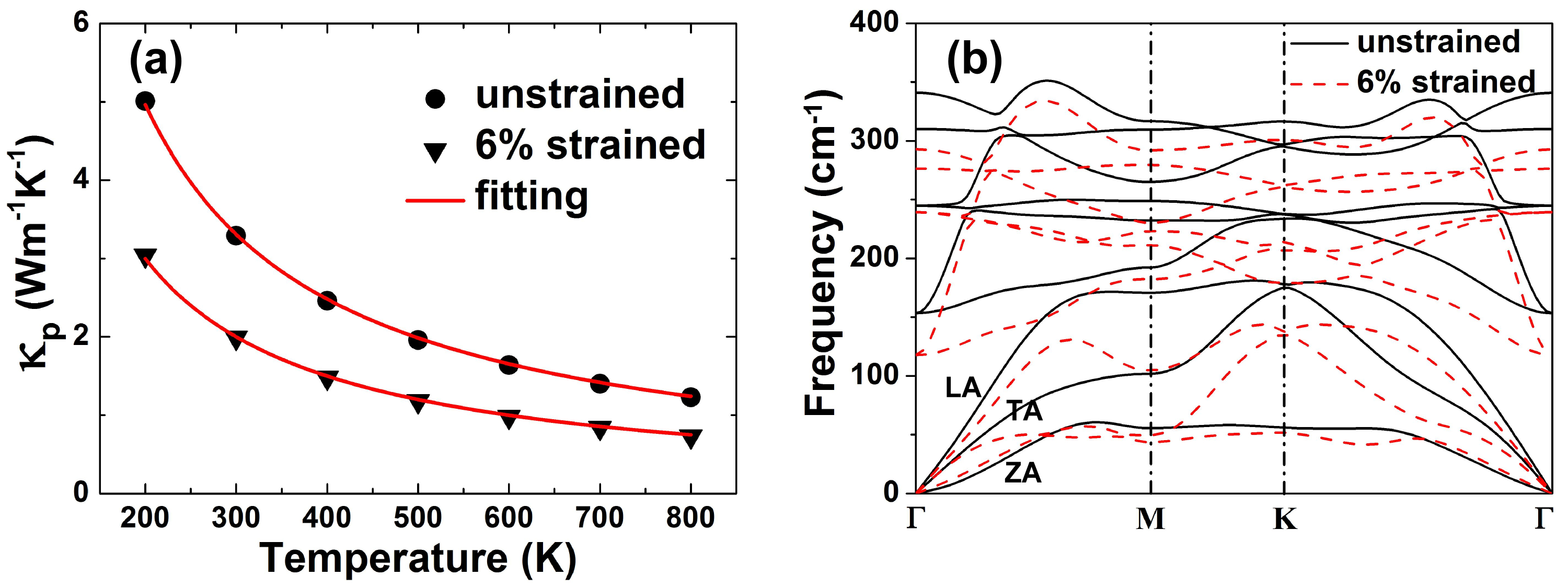}\caption{\label{fig4-phonon}(a) Temperature dependence of the lattice thermal conductivities and (b) phonon dispersions of the unstrained and 6\% strained ZrS$_{2}$ monolayers.}
\end{figure*}

As mentioned above, within our method, the electrical conductivity $\sigma$ can only be calculated with the relaxation time $\tau$ inserted. The relaxation time is determined by
\begin{equation}\label{1}
   \mu=e\tau/m^*,
\end{equation}
where $\mu$ is the carrier mobility and $m^*$ is the effective mass. The carrier mobility $\mu$ of the ZrS$_{2}$ monolayer can be calculated using the deformation potential (DP) theory based on the effective mass approximation:\cite{DP1,DP2,DP3}
\begin{equation}\label{2}
    \mu=\frac{e\hbar^3C}{k_BTm^*m_dE_1^2},
\end{equation}
where $m_d$ is the average effective mass defined by $m_d=\sqrt{m^*_xm^*_y}$. $C$ is the elastic modulus and can be determined by $C=(\partial^2 E/\partial\delta^2)/S_0$, where $E$, $\delta$, $V_0$, and $S_0$ are, respectively, the total energy, the applied strain, the volume, and the area of the investigated system. The DP constant $E_1$ is obtained by $E_1=dE_{edge}/d\delta$, where $\delta$ is the applied strain by a step of 0.5\% and $E_{edge}$ is the energy of the band edges (VBM for the holes and CBM for the electrons). Since the ZrS$_{2}$ monolayer has the largest power factor at the strain of 6\%, we will only consider this strain condition. The calculated $m^*$, $m_d$, $C$, $E_1$, and room-temperature $\mu$ and relaxation time $\tau$ of the unstrained and 6\% strained ZrS$_{2}$ monolayers are summarized in Table I. For the unstrained case, the effective mass of the electron along the armchair direction is much larger than that along the zigzag direction, since the conduction band dispersion around the $M$ point is highly anisotropic (see Fig. 1(c)), that is, the conduction band is much flatter along the $\Gamma$-$M$ direction (armchair direction in real space) than that along the $M$-$K$ direction (zigzag direction in the real space). However, the valence band dispersion around the $\Gamma$ point is nearly isotropic, which results in the almost same effective masses of the hole along the zigzag and armchair directions. The case for the strained ZrS$_{2}$ monolayer is similar, except that the corresponding effective masses are enhanced by the strain. The elastic modulus $C$ and DP constant $E_1$ for both the electron and hole have little difference along the two directions. Accordingly, the anisotropy of the carrier mobilities is dominated by the corresponding anisotropy of the carrier effective masses. For the electron, the mobility along the zigzag direction is much larger than that along the armchair direction. The electron mobility along the zigzag direction of the unstrained system is 1045.3 cm$^2$V$^{-1}$s$^{-1}$, comparing favorably with that of the MoS$_{2}$ monolayer.\cite{MoS2-monolayer} The applied strain reduces both the electron and hole mobilities, mainly due to the increased effective masses. Based on Eqs. (1) and (2), the relaxation time $\tau$ will depend on the average effective mass $m_d$ but not $m^*$, so the anisotropy of $m^*$ will not be reflected in $\tau$. As a result, there is little difference in the relaxation time $\tau$ along different directions. In the following, we will use the averaged $\tau$ of the two directions to estimate the TE performance.

Next, we will investigate the strain effect on the thermal transport properties of the ZrS$_{2}$ monolayer. The lattice thermal conductivities $\kappa_p$ of the unstrained and 6\% strained ZrS$_{2}$ monolayers are shown in Fig. 4(a). For both cases, the $\kappa_p$ decreases as increasing the temperature, following a $T^{-1}$ dependence, as indicated by the fitting lines in the figure. This implies the dominant scattering mechanism is the Umklapp process. Moreover, in the temperature range investigated, the thermal conductivities of the 6\% strained ZrS$_{2}$ monolayer are much smaller than those of the unstrained system. In particular, at 300 K, the $\kappa_p$ decreases from 3.29 to 1.99 W/(mK), reduced by 40\% when the strain of 6\% is applied. To investigate the origin of the reduction in the thermal conductivity, in Fig. 4(b), we plot the corresponding phonon dispersions. At the strain of 6\%, the phonon dispersions of transverse and longitudinal acoustic (TA and LA) modes become softened, while the out-of-plane acoustic (ZA) mode is slightly stiffened. The contribution of each phonon mode to the total thermal conductivity can be expressed as
\begin{equation}\label{3}
    \kappa_i({\bf{q}})=C_i({\bf{q}})\upsilon_i^2({\bf{q}})\tau_i({\bf{q}}),
\end{equation}
where $C_i$, $\upsilon_i$, and $\tau_i$ are the specific heat, group velocity and phonon relaxation time, respectively; ${\bf{q}}$ is the wave vector. The group velocity is calculated based on $\upsilon_i({\bf{q}})=\frac{\partial\omega}{\partial {\bf{q}}}$. Our results show that the three acoustic phonon branches contribute mostly to the thermal conductivity. The phonon softening leads to the reduced group velocity, which in turn reduces the $\kappa_p$. Moreover, because of the phonon softening, the phonons with fixed number of frequencies gather in a narrower frequency range, which increases the chance of the phonon scattering and thus the phonon relaxation time decreases. This is an additional source of the reduction in the thermal conductivity. In contrast, the phonon stiffening often increases the thermal conductivity. The calculated $\kappa_p$ of each mode confirms the above analysis, that is, the lattice thermal conductivity from the contribution of the ZA mode is indeed increased, while those from the TA and LA modes are reduced. However, the reduction originating from the softening of the TA and LA modes predominates, so the total lattice thermal conductivity is reduced by the strain.

Combining the electronic and thermal transport properties, we now evaluate the TE performance of the ZrS$_{2}$ monolayer. The room-temperature $ZT$ values of the unstrained and 6\% strained systems are shown in Fig. 5. For the unstrained ZrS$_{2}$ monolayer, the maximal $ZT$ value of the $n$-type doped system is much larger than that of the $p$-type doped one. When the strain of 6\% is applied, the maximal $ZT$ value of 2.40 is obtained for the $p$-type doping, which is 4.3 times larger than that of the unstrained system. For the $n$-type doping, the enhancement of the maximal $ZT$ value is relatively smaller, from 1.65 for the unstrained system to 1.76 for the 6\% strained one. The increase of the Seebeck coefficient as well as the decrease of the lattice thermal conductivity contribute to the enhanced TE performance.

\begin{figure}
\includegraphics[width=0.9\columnwidth]{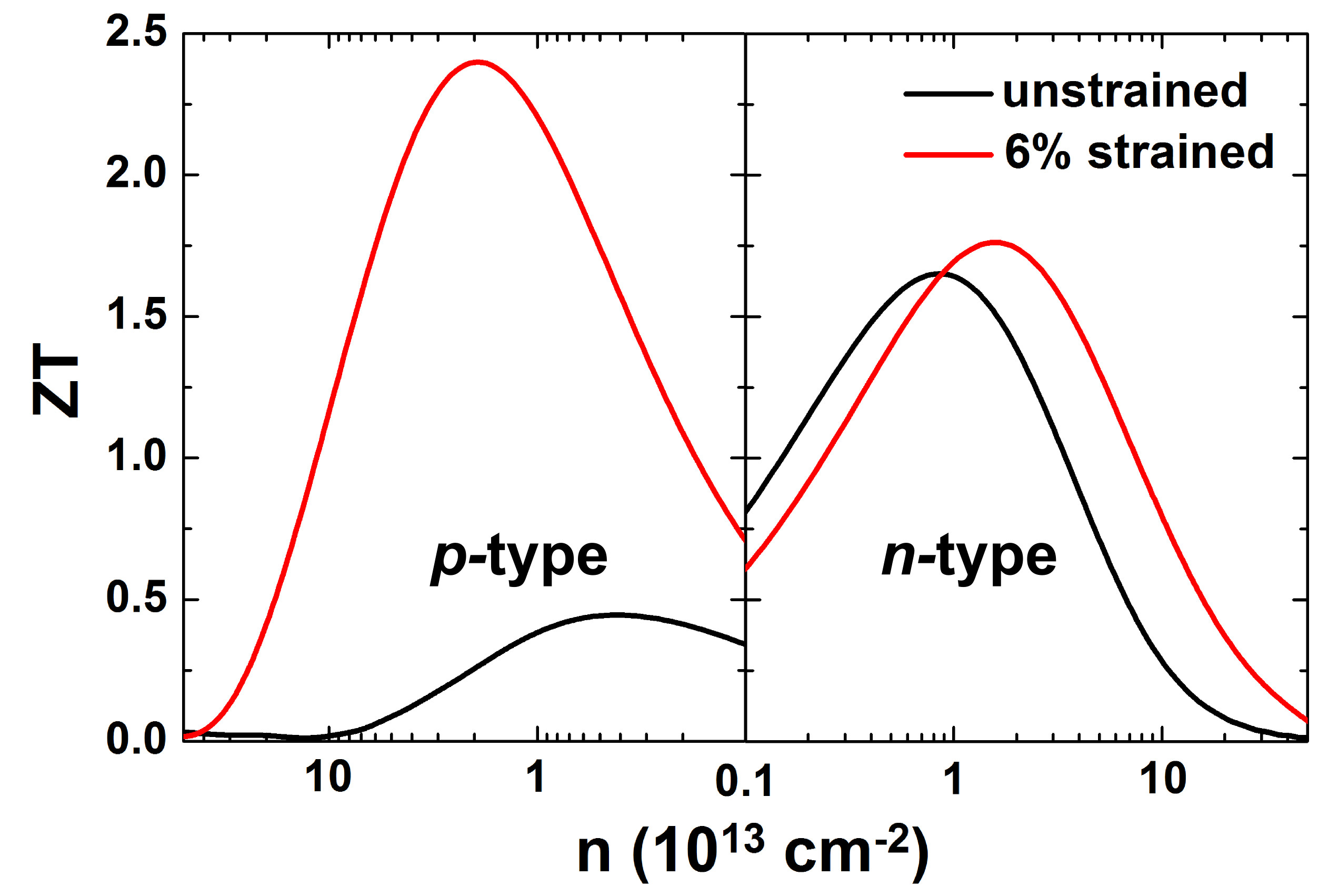}\caption{\label{fig5-ZT} $ZT$ values at 300 K as a function of the carrier concentration of the unstrained and 6\% strained ZrS$_{2}$ monolayers.}
\end{figure}

In summary, we have investigated the strain effect on the electronic, phonon, and TE properties of the ZrS$_{2}$ monolayer. The band gap first increases and then decreases as increasing the biaxial strain, reaching the maximum at the strain of 6\%. The Seebeck coefficient of the ZrS$_{2}$ monolayer is found to be effectively tuned by the strain $via$ the band-gap engineering. The increase of the Seebeck coefficient compensates the decrease of the electrical conductivity and thus the maximal power factor of the ZrS$_{2}$ monolayer increases as increasing the band gap. At the strain of 6\%, the TE performance of the ZrS$_{2}$ monolayer is greatly enhanced, due to the simultaneous increase of the Seebeck coefficient and decrease of the thermal conductivity. Our results show that the applied strain is an efficient method to enhance the TE performance of the ZrS$_{2}$ monolayer.

This work was supported by the National Natural Science Foundation of China under Contracts No. 11274311, 11404340, and No. U1232139, the Anhui Provincial Natural Science Foundation under Contract No. 1408085MA11, the China Postdoctoral Science Foundations (Grant Nos. 2014M550352 and 2015T80670). The calculation was partially performed at the Center for Computational Science, CASHIPS.

\end{document}